\documentclass[aps,prx,reprint,superscriptaddress,nofootinbib,longbibliography]{revtex4-2}

\usepackage{amsmath,amssymb,mathtools,bm}
\usepackage{amsthm}
\usepackage{booktabs}
\usepackage{multirow}
\usepackage{longtable}
\usepackage{array}
\usepackage[version=4]{mhchem}
\usepackage[]{hyperref}
\usepackage{algorithmicx}
\usepackage[noend]{algpseudocode}
\usepackage{tikz}

\usepackage{xr}
\newcommand{\epsM}{\varepsilon^2 M}
\newcommand{\Var}{\operatorname{Var}}
\newcommand{\Cov}{\operatorname{Cov}}

\newcommand{\calG}{\mathcal{G}}

\begin{document}

\title{Reducing quantum measurements in qubit-based overlapping grouping methods for quantum energy estimation through better initializations}

\author{Isaac L. Huidobro-Meezs}
\email{huidobri@mcmaster.ca}
\affiliation{Department of Chemistry and Chemical Biology, McMaster University,\\ Hamilton, ON L8S 4M1, Canada}

\author{Rodrigo A. Vargas-Hern\'andez}
\email{vargashr@mcmaster.ca}
\affiliation{Department of Chemistry and Chemical Biology, McMaster University,\\ Hamilton, ON L8S 4M1, Canada}
\affiliation{Brockhouse Institute for Materials Research, McMaster University,\\ Hamilton, ON, Canada}

\date{\today}

\begin{abstract}
The measurement cost for estimating expectation values of Hamiltonians is a central bottleneck in variational quantum algorithms. Grouping strategies significantly reduce this cost, with overlapping techniques as the state of the art. Overlapping grouping methods require i) a non-overlapping grouping of the Hamiltonian, typically obtained from the Sorted Insertion (SI) algorithm as initialization, and ii) the construction of covariance dictionaries from approximate wavefunctions to guide the optimization. It was recently shown that different initializations can potentially reduce measurement costs for overlapping methods. Motivated by these findings, we introduce variance-aware SI (VarSI), a family of covariance-informed non-overlapping Pauli grouping heuristics to reduce measurement counts. VarSI grouping leverages the covariance dictionaries, already required by overlapping methods, to construct better non-overlapping groups. We propose three variants:  a global greedy grouping insertion rule, a variance-informed SI analog, and a local refinement step initialized from SI or our variance-informed variant. We showcase the use of the VarSI groupings as initialization of overlapping methods using the iterative coefficient-splitting (ICS) algorithm. Molecular benchmarks with 130 Hamiltonians demonstrate consistent, non-overlapping measurement improvements over SI of 36--39\% and enhanced downstream ICS results when initialized from VarSI groups. We find that the initializations considered here achieve larger reductions of up to 48\% for systems whose covariances deviate from idealized cases. Compared to the standard SI initialization, VarSI reaches mean reductions of 10.3--13.5\% depending on qubit mappings and covariance dictionaries used. These results show that non-overlapping grouping remains a consequential design step even when the final estimator uses overlapping fragments.
\end{abstract}

\maketitle

\section{Introduction}
\label{sec:intro}
Several variational quantum algorithms for problems like energy estimation of many-body Hamiltonians, including electronic structure problems \cite{VQARev,Peruzzo2014VQE,VarExcited,ADAPT-VQE,ContractedSchrodinger,QSE,KrylovSubspace,McClean2016Theory,Preskill2018NISQ,IncreasingAccuracyNoExtraQuant,QuantumPowerMethod,Q-EOM}, and quantum machine learning (QML) protocols \cite{QuantumBoltzmannML,QuantumCircuitLearning,MeasurementQuantumNN,VariationalQuantumGenModeling,VarQuantumHamEngineering,DeepQuantumNNs,RegressionQuantum}, ultimately require extracting expectation values of observables from measurement outcomes on a quantum device. For these variational algorithms, the Hamiltonian or a generic operator is mapped to a qubit operator expanded as a linear combination,
\begin{equation}
    H = \sum_{k=1}^{N_P} c_k P_k;  \ \ P_k = \bigotimes_{n = 1}^{N_q}\sigma_n^{(k)},
    \label{eq:hamiltonian}
\end{equation}
where $P_k$ are Pauli words that correspond to the tensor product of Pauli operators and identities for each qubit $\sigma_n^{(k)} \in \{\hat{x}_n, \hat{y}_n, \hat{z}_n, \hat{1}_n \}$ associated to the $k$-th Pauli word and $c_k$ the coefficients of the linear combination. For quantum chemistry applications, the number of Pauli words in molecular Hamiltonians scales $\mathcal{O}(N_q^4)$ \cite{10.1063/1.4768229} with the number of qubits/spin-orbitals ($N_q$), making the direct measurement of each Pauli operator required to obtain the necessary expectation values impractical \cite{MeasurementsRoadblock,Patel2025Review}.

A widely used alternative for the measurement problem is partitioning the Pauli words of the Hamiltonian into compatible groups,
\begin{equation}
    H = c_0 I + \sum_{\alpha=1}^{N_f} H_\alpha, \label{eq:partition}
\end{equation}
where $N_f$ is the total number of groups, with a given grouping defined as $\mathcal{G} = \{G_\alpha\}_{i=1}^{N_f}$. In Eq.~\eqref{eq:partition}, each group $\alpha$ forms a subhamiltonian, 
\begin{equation}
    H_\alpha = \sum_{i\in G_\alpha} c_i P_i, \label{eq:groups}
\end{equation}
so that all the Pauli words in $G_\alpha$ can be measured in a common basis \cite{Verteletskyi2020MinimumClique,Patel2025Review,Wu2023OverlappedGrouping}. 
Two compatibility notions are especially common: Qubit-wise commutativity (QWC) and full commutativity (FC). QWC allows tensor-product-basis measurements using only local Clifford rotations. In contrast, FC permits larger compatible fragments and lower measurement counts at the cost of more general Clifford rotations \cite{Jena2019PauliPartitioning,Gokhale2020ON3,Yen2020AllCompatible,Crawford2021SortedInsertion,Yen2023Deterministic,Patel2025Review}. Graph-based formulations of the grouping problem connect QWC or FC partitioning to clique-cover-type problems, and exact optimization is replaced, in practice, by heuristic algorithms \cite{Verteletskyi2020MinimumClique,Crawford2021SortedInsertion}.

Once a set of groups is defined, the optimal distribution of measurements can be found from the variances of each group, leading to the number of measurements required to reach a target standard error $\varepsilon$ being defined as,
\begin{equation}
    \epsM(\calG)
    =\left(\sum_{\alpha=1}^{N_f}\sqrt{\Var_\psi(H_\alpha)}\right)^2,
    \label{eq:intro-objective}
\end{equation}
where $\calG=\{G_\alpha\}$ is the given grouping and $\psi$ is the wavefunction implemented on the quantum device \cite{Yen2023Deterministic}.  
Sorted insertion (SI), a greedy policy algorithm, orders Pauli words by coefficient magnitude before greedy insertion. SI is designed to reduce this metric, Eq.~\eqref{eq:intro-objective}, and can outperform conventional graph-coloring heuristics designed only to reduce the number of fragments, for which it has been widely used \cite{Crawford2021SortedInsertion}. Although this partitioning problem should also account for the associated circuit implementation costs \cite{huidobromeezs2025discreteflowbasedgenerativemodels,k-commutativity}, the number of measurements remains a central resource metric for Hamiltonian estimation, even in idealized error-free settings \cite{MeasurementsRoadblock}, especially when the additional basis-change circuits are Clifford-dominated and therefore comparatively inexpensive relative to non-Clifford fault-tolerant resources \cite{Campbell2017,OptimalHadamardForSynthesis}.

Measurement reduction has expanded in several complementary directions. For example, Fermionic methods exploit the structure of molecular Hamiltonians, looking at partitions that can be diagonalized into an Ising form directly on the fermionic space \cite{Choi2023Fluid}, while classical-shadow variants replace deterministic bases by randomized or derandomized measurement ensembles \cite{Huggins2021NoiseResilient,Huang2021Derandomization,Hadfield2022LocallyBiased}. 
Within qubit-based partitioning approaches, the state-of-the-art methods go beyond non-overlapping partitions. Overlapped grouping measurements, such as iterative coefficient splitting (ICS) \cite{Yen2023Deterministic} and Shared Pauli Products (SPP) \cite{Choi2022Ghost}, exploit the fact that a Pauli word compatible with multiple fragments can contribute information in multiple measurement contexts \cite{Wu2023OverlappedGrouping}.
In ICS, a Pauli word compatible with multiple fragments can be assigned coefficients related to each group $\alpha$, $c_i^{(\alpha)}$, satisfying
\begin{equation}
    c_i=\sum_{\alpha\in I_i}c_i^{(\alpha)},
\end{equation}
where $I_i$ is the set of fragments in which $P_i$ is measurable. ICS optimizes these coefficients $c_i^{(\alpha)}$, together with shot allocations, using covariance estimates \cite{Yen2023Deterministic}. SPP instead introduces Pauli words with a net zero coefficient in the total Hamiltonian that can be added across fragments to change fragment variances without changing the target expectation value \cite{Choi2022Ghost,Patel2025Review}. Both methods employ covariance dictionaries constructed from approximate wavefunctions as input to the optimization protocol specific to each method. 
The required covariance dictionaries can also be dynamically constructed and continuously improved directly from the initial measurement results of the quantum device \cite{Shlosberg2023adaptiveestimation}, allowing for a more realistic application of these measurement allocation techniques, since their performance depends on the implemented wavefunction. 

Approaches like ICS and SPP can substantially reduce estimator variance, but they still rely on a set of compatible seed fragments for initialization, often obtained via SI, which we denote as SI-ICS in this work.
Moreover, recent results employing generative models \cite{HuidobroMeezs2024GFlowNets,huidobromeezs2025discreteflowbasedgenerativemodels}, trained to reduce measurement counts and hardware resources via a reward-driven stochastic grouping policy, have shown that different initializations of these methods can further reduce measurement costs. 
A similar result has been found in fermionic space, where greedy full-rank optimization of initial fragments outperforms cheap low-rank optimization in fluid-fermionic fragments \cite{Choi2023Fluid}. 
Improved non-overlapping Pauli group initialization may transfer to any Pauli-sum observable with nontrivial commutation structure, as measurement grouping techniques are not restricted to molecular Hamiltonians. This strategy can be employed to estimate quantum observables expressed as Pauli sums in generic variational quantum algorithms, including applications towards many-body/lattice Hamiltonians \cite{PhysRevB.102.075104,PhysRevB.76.180407,PhysRevD.108.094513,Zhang2023simulatinggauge}, and QML \cite{QuantumBoltzmannML,QuantumCircuitLearning,MeasurementQuantumNN,VariationalQuantumGenModeling,VarQuantumHamEngineering,DeepQuantumNNs,RegressionQuantum}.
The quality of a non-overlapping grouping, therefore, remains a consequential step towards reducing measurement costs and is a clear area for improvement in qubit-based methods.

In this paper, we present variance-aware sorted insertion (\textbf{VarSI}), a family of covariance-informed heuristic algorithms for generating non-overlapping compatible Pauli groups, along with greedy refinement steps. VarSI employs a metric directly related to the objective in Eq.~\eqref{eq:intro-objective}, using a dictionary of Pauli covariances that is already necessary for overlapping methods, incurring no additional cost.  We make the following contributions: first, we derive the covariance update that allows scoring of candidate insertions.  Second, we define and compare a global greedy variance-based allocation method (\textbf{VarSI-G}), an ordered direct analog to sorted insertion (\textbf{VarSI-O}), and refinement variants for both SI and VarSI-O (\textbf{VarSI-R}, \textbf{VarSI-OR}) against standard SI. The refinement variant is guaranteed to decrease $\epsM$ and therefore returns a grouping no worse than the starting grouping. We show consistent improvements over the SI baselines for VarSI-O, VarSI-R, and VarSI-OR. Finally, we use these heuristic algorithms as initializations for ICS and test them on various molecular benchmarks, including comparisons with the more expensive SPP method and tests on strongly correlated systems, using both approximate and exact covariance dictionaries and different fermion-to-qubit mappings.

\section{Methods}
\subsection{Measurement objective and covariance form}
\label{sec:objective}
For a fixed grouping $\calG$, suppose that fragment $H_\alpha$ is measured with $m_\alpha$ independent shots. The estimator error for the total Hamiltonian is given by \cite{OrigErrorExpression,Romero_2019,Rubin_2018},
\begin{equation}
    \varepsilon^2=\sum_{\alpha=1}^{N_f}\frac{V_\alpha}{m_\alpha};
    \quad V_\alpha=\Var_\psi(H_\alpha),
    \label{eq:estimator-var}
\end{equation}
 with the total number of measurements given by $M =\sum_{\alpha=1}^{N_f}m_\alpha$. Optimizing the $m_\alpha$ coefficients yields Eq.~\eqref{eq:intro-objective} \cite{Crawford2021SortedInsertion,Yen2023Deterministic}. We consider the effect on the metric, Eq. \eqref{eq:intro-objective}, of the addition of an element to a group $G_\alpha$. We define the state-dependent covariance matrix of unit-coefficient Pauli words as,
\begin{equation}
    C_{ij}=\Cov_\psi(P_i,P_j)
    =\langle P_iP_j\rangle_\psi
     -\langle P_i\rangle_\psi\langle P_j\rangle_\psi .
    \label{eq:cov-def}
\end{equation}
For a fragment $G_\alpha$,
\begin{equation}
    V_{G_\alpha}=\Var_\psi\left(\sum_{i\in G_\alpha}c_iP_i\right)
    =\sum_{i,j\in G_\alpha}c_ic_jC_{ij}.
    \label{eq:group-var}
\end{equation}
If the term $c_kP_k$ is added to $G_\alpha$, then the variance of the group is modified as,
\begin{align} 
    V_{G_\alpha\cup k}
    &=V_{G_\alpha}+c_k^2C_{kk}
      +2c_k\sum_{i\in G_\alpha}c_iC_{ik},\label{eq:addition}
\end{align}
where the second term $c_k^2C_{kk}$ is the variance of the new term, and the third term, $2c_k\sum_{i\in G_\alpha}c_iC_{ik}$, is the covariance of the new element with the elements already existing in $G_\alpha$.
This equation, Eq.~\eqref{eq:addition}, will serve as the basis for the scoring rule used across all VarSI variants presented here. 

From Eq.~\eqref{eq:addition}, one can identify an extension that further reduces the classical cost of variance-based measurement allocation methods.
One can employ data-driven models to generate approximate initializations for the covariance dictionaries by selecting families of wavefunctions and scaling the dictionaries based on the Hamiltonian's coefficients.
For example, these dictionaries can be initialized with pre-trained coupled-cluster wavefunctions \cite{Mole}, approximations based on the 1- and 2-body reduced density matrices \cite{ML1RDM,b2026ML2RDM}, or other inexpensive generative policies that achieve performance comparable to classical methods at lower cost \cite{FlowVQE}. Although such extensions are beyond the scope of this work, we expect them to enter the field soon.
\begin{table}[t!]
\caption{Worst-case grouping costs with a precomputed covariance dictionary.  $N$ is the number of Pauli words, $n_q$ is the number of qubits, and $N_S$ is the number of refinement sweeps.}
\label{tab:costs}
\begin{tabular}{ll}
\hline
Method & Worst-case grouping cost \\ 
\hline
SI & $O(N\log N+N^2n_q)$ \\
VarSI-O & $O(N\log N+N^2(n_q+1))$ \\
VarSI-G & $O(N^3(n_q+1))$ \\
VarSI-R, naive & $O(N_S(N^2n_q+N^3))$ \\
VarSI-R, cached & $O(N_SN^2(n_q+1))$ \\ \hline
\end{tabular}
\end{table}

\begin{table*}[t]
\caption{$\epsM$ for STO-3G Hamiltonians under FC. The VarSI groupings were obtained using CISD covariances, and the reported final variances employ the exact wavefunctions. \ce{BeH2}$^{(s)}$ and \ce{H2O}$^{(s)}$ use an interatomic distance of 3.0 and 2.2\AA, respectively. Columns containing $S=(100/500)$ report results obtained with 100 and 500 refinement sweeps. Bold and underlined entries indicate the \textbf{best} and \underline{second-best} displayed values, respectively, within each grouping type. SPP and so-SPP values taken from Ref. \cite{Choi2022Ghost}.}

\label{tab:main-spp-comparison}
\fontsize{7.0pt}{7.1pt}\selectfont
\setlength{\tabcolsep}{1.0pt}
\renewcommand{\arraystretch}{1.12}
\begin{tabular*}{\textwidth}{@{\extracolsep{\fill}}lcc@{\hspace{0.1em}}ccccc@{\hspace{0.4em}}!{\vrule width 0.5pt}@{\hspace{0.35em}}ccccccc@{}}
\toprule
& & & \multicolumn{5}{c}{\textbf{Non-overlapping groupings}} & \multicolumn{7}{c}{\textbf{Overlapping methods}} \\
\cmidrule(lr){4-8}\cmidrule(l){9-15}
System & $N_q$ & $N_P$ & SI & VarSI-G & VarSI-O & \shortstack{VarSI-R;\\$S=(100/500)$} & \shortstack{VarSI-OR;\\$S=(100/500)$} & \shortstack{SI\\-ICS} & \shortstack{VarSI-G\\-ICS} & \shortstack{VarSI-O\\-ICS} & \shortstack{VarSI-R-ICS;\\$S=(100/500)$} & \shortstack{VarSI-OR-ICS;\\$S=(100/500)$} & SPP & so-SPP \\
\midrule \vspace{0.1cm}
& & \multicolumn{12}{c}{{\normalsize Jordan--Wigner}} & \\
\ce{LiH} & 12 & 630 & 0.882 & 0.970 & 0.580 & \underline{0.313}/\textbf{0.312} & 0.380/0.380 & 0.234 & 0.191 & 0.182 & 0.230/0.230 & \underline{0.140}/\textbf{0.140} & 0.158 & 0.172 \\
\ce{BeH2} & 14 & 665 & 1.111 & 2.466 & 1.120 & 0.731/0.731 & \underline{0.704}/\textbf{0.704} & 0.482 & 0.684 & 0.438 & 0.394/0.394 & \underline{0.354}/\textbf{0.354} & 0.370 & 0.413 \\
\ce{BeH2}$^{(s)}$ & 14 & 665 & 1.994 & 4.601 & 1.872 & \underline{1.647}/\textbf{1.647} & 1.661/1.661 & 2.267 & 1.252 & 1.715 & 2.276/2.276 & \underline{1.233}/1.233 & \textbf{0.928} & 1.470 \\
\ce{H2O} & 14 & 1085 & 7.589 & 31.676 & 11.253 & \underline{3.438}/\textbf{3.438} & 4.817/4.764 & 1.521 & 2.315 & 1.365 & \underline{1.166}/\textbf{1.166} & 1.372/1.359 & 1.330 & 1.540 \\
\ce{H2O}$^{(s)}$ & 14 & 1085 & 3.659 & 25.862 & 3.284 & \underline{2.848}/\textbf{2.838} & 3.163/3.163 & 0.961 & 2.701 & 1.217 & \underline{0.773}/\textbf{0.767} & 1.174/1.174 & 1.050 & 1.730 \\
\ce{NH3} & 16 & 3608 & 18.751 & 42.406 & 13.831 & 7.814/\underline{6.569} & 7.135/\textbf{6.225} & 3.424 & 6.412 & 2.646 & 3.243/2.875 & 2.553/\underline{2.436} & \textbf{2.240} & 2.900 \\
\midrule \vspace{0.1cm}
& & \multicolumn{12}{c}{{\normalsize Bravyi--Kitaev}} & \\
\ce{LiH} & 12 & 630 & 0.882 & 0.970 & 0.580 & \underline{0.313}/\textbf{0.312} & 0.399/0.399 & 0.234 & 0.191 & 0.182 & 0.230/0.230 & \underline{0.141}/\textbf{0.141} & 0.155 & 0.169 \\
\ce{BeH2} & 14 & 665 & 1.111 & 2.587 & 1.120 & \underline{0.731}/\textbf{0.731} & 0.734/0.734 & 0.482 & 0.551 & 0.413 & 0.394/0.394 & 0.379/\underline{0.379} & \textbf{0.360} & 0.406 \\
\ce{BeH2}$^{(s)}$ & 14 & 665 & 1.994 & 5.343 & 1.874 & \underline{1.647}/\textbf{1.647} & 1.712/1.712 & 2.267 & \underline{1.021} & 1.749 & 2.275/2.275 & 1.591/1.591 & \textbf{0.937} & 1.610 \\
\ce{H2O} & 14 & 1085 & 7.589 & 31.676 & 11.188 & \underline{3.438}/\textbf{3.438} & 4.900/4.853 & 1.521 & 2.315 & 1.403 & \underline{1.166}/\textbf{1.166} & 1.375/1.387 & 1.340 & 1.540 \\
\ce{H2O}$^{(s)}$ & 14 & 1085 & 3.659 & 45.273 & 3.284 & \underline{2.848}/\textbf{2.838} & 3.157/3.157 & 0.961 & 2.121 & 1.209 & \underline{0.773}/\textbf{0.767} & 1.171/1.171 & 1.030 & 1.720 \\
\ce{NH3} & 16 & 3608 & 18.751 & 47.375 & 14.251 & 7.814/\textbf{6.569} & 7.527/\underline{6.707} & 3.424 & 6.670 & 2.914 & 3.243/2.875 & 2.810/\underline{2.665} & \textbf{2.230} & 2.890 \\
\midrule
\multicolumn{3}{l}{Mean reduction (\%)} & -- & -269.3 & 4.5 & 41.9/43.1 & 36.2/37.1 & -- & -42.3 & 9.9 & 11.3/13.2 & 18.5/19.1 & 25.8 & 1.5 \\
\multicolumn{3}{l}{Max reduction (\%)} & -- & -10.0 & 34.2 & 64.5/65.0 & 61.9/66.8 & -- & 55.0 & 24.3 & 23.3/23.3 & 45.6/45.6 & 59.1 & 35.2 \\
\bottomrule
\end{tabular*}
\end{table*}

\subsection{Variance-aware sorted insertion algorithms}
\label{sec:methods}
All algorithms considered in this work return non-overlapping groups and enforce the chosen compatibility rule, FC or QWC, by allowing a term to enter a group only if it is compatible with every term already in that group. We limit this study to FC groupings as they provide the lowest measurement counts \cite{Patel2025Review,Choi2022Ghost,Yen2023Deterministic}; however, VarSI can be employed with QWC and k-commuting groups \cite{DalFavero2024KCommutativity}. We define the total score for a grouping $\calG$ as,
\begin{equation}
    S(\calG)=\sum_{G\in\calG}\sqrt{V_G},
    \qquad
    \epsM(\calG)=S(\calG)^2.
\end{equation}
Since squaring is monotone for $S\geq 0$, each local decision can be made by minimizing either $S$ or $\epsM$ from Eq. \ref{eq:intro-objective}.

We report the computational cost of the proposed algorithms in Table~\ref{tab:costs}. Please refer to Sections I and II of the supporting material (SM) for a detailed explanation of the computational cost and pseudocode of this and the other VarSI algorithms.\\

\textbf{Ordered VarSI} (\textbf{VarSI-O}) is the closest analog of the standard SI algorithm and the main variant we introduce due to its computational scaling for Hamiltonians composed of a large number of Pauli terms. SI orders terms by $|c_i|$, and then inserts each Pauli term into the first compatible group. Instead, VarSI-O orders terms by their single-term variance contribution,
\begin{equation}
    v_i=c_i^2C_{ii},
\end{equation}
and then processes this fixed list once. For the current term $k$, each compatible existing group $G_\alpha$ is scored by
\begin{equation}
    S_{\alpha,k}
    =S(\calG)-\sqrt{V_{G_\alpha}}
    +\sqrt{V_{G_\alpha\cup k}},
    \label{eq:ordered-score}
\end{equation}
where $V_{G_\alpha\cup k}$ is evaluated using Eq.~\eqref{eq:addition}.  The term is inserted into the compatible group with the smallest $S_{\alpha,k}$.  If there is no compatible group, a new singleton group is opened.\\

\textbf{Global greedy VarSI} (\textbf{VarSI-G}) starts with the largest single-term variance, then scans all remaining terms and compatible destinations. The accepted move is the term-group insertion that gives the lowest immediate value of $\epsM$. If a term has no compatible destination, opening a singleton group is considered for that term. VarSI-G therefore has more freedom than VarSI-O because it can dynamically change the insertion order. This freedom is not always beneficial: a low-cost insertion can be selected because it has little immediate variance effect, even if it later prevents a higher-variance term from entering a better fragment.\\

\textbf{Refinement from a given grouping}. As greedy algorithms cannot guarantee optimality, we consider an additional refinement step. \textbf{VarSI-R} and \textbf{VarSI-OR} are local refinements initialized from SI and VarSI-O groupings, respectively. Starting from $\calG_0=\calG_\mathrm{SI/VarSI-O}$, the algorithm considers moving one term from its current group to another compatible group. A sweep evaluates all such one-term relocations and accepts the move with the lowest candidate objective only if the candidate objective is strictly smaller than the current objective. The process terminates when no improving move exists or when a specified maximum number of sweeps is reached.

\section{Results and Discussion} \label{sec:results}

In Table \ref{tab:main-spp-comparison}, we show the comparison of $\epsM$ for a series of molecular systems, \ce{LiH}, \ce{BeH2}, \ce{H2O}, and \ce{NH3}. The qubit Hamiltonians were obtained using the Jordan--Wigner (JW) \cite{JordanWigner} and Bravyi--Kitaev (BK) \cite{BK,10.1063/1.4768229} transformations of the fermionic Hamiltonians in the STO-3G basis set using Tequila \cite{Kottmann2021Tequila} and PySCF \cite{PySCF,PySCF2} with a $1\,\text{\AA}$ \ce{X-H} distance for all of them and angles $\angle\mathrm{H}\mathrm{Be}\mathrm{H}=180^\circ$ (for \ce{BeH2}), $\angle\mathrm{H}\mathrm{O}\mathrm{H}=107.6^\circ$ (for \ce{H2O}), and $\angle\mathrm{H}\mathrm{N}\mathrm{H}=107^\circ$ (for \ce{NH3}). For \ce{BeH2} and \ce{H2O}, we also considered stretched configurations, labeled by $s$ as a superscript, with interatomic distances of $3\,\text{\AA}$ and $2.2\,\text{\AA}$, respectively. These molecular systems were also studied in Ref. \cite{Choi2022Ghost} for SPP and its cheaper sequentially optimized version (so-SPP). We initialize ICS with the SI groupings and the VarSI-G/O/R/OR groupings, and refer to the resulting groupings as SI-ICS and VarSI-G/O/R/OR-ICS, respectively. For the results in Table \ref{tab:main-spp-comparison}, we used the CISD wavefunction to construct the covariance dictionaries and report the $\epsM$ result with the exact ground-state wavefunction obtained from diagonalizing the qubit Hamiltonian. The proposed methods here, with the exception of VarSI-G, reduce the number of measurements compared to the standard SI baseline by an average of 36--43\% for the refinement variants, even while using an approximate wavefunction for the scoring steps.\\

For the overlapping results in Table \ref{tab:main-spp-comparison}, our results show that the VarSI policies allow the cheaper ICS method to compete with the SPP method. Importantly, our method retains the scaling of ICS but has a larger constant factor than the standard SI initialization; for further details, see SM. VarSI can be easily implemented as an extension that interfaces with existing software packages, such as Tequila \cite{Kottmann2021Tequila}, Pennylane \cite{pennylane}, and OpenFermion \cite{McClean2020OpenFermion}, as shown in the included example in the \href{https://github.com/ishume94/VarSI}{repository}. We found no significant improvement with increasing the number of sweep steps, so we limited subsequent studies to 100 sweep steps. From this small benchmark, we found that the VarSI-O-ICS, VarSI-R-ICS, and VarSI-OR-ICS variants consistently reduce measurement costs, with VarSI-OR-ICS reducing measurement costs by an average of 18.5\% with 100 sweeps and 19.1\% with 500 sweeps relative to the commonly used SI-ICS method.\\

We extend our study to a larger set of molecular Hamiltonians to assess the improvement over standard SI-ICS that can be obtained from the better initializations proposed in the present paper. We explored first a series of molecular Hamiltonians inspired by \texttt{HamLib} \cite{HamLib}. To maintain operator consistency, we regenerated the Hamiltonians using Tequila \cite{Kottmann2021Tequila} with PySCF \cite{PySCF,PySCF2} as the backend to retain all-electron operators and options, including frozen-core orbitals. For these Hamiltonians, we considered covariance dictionaries built from the ground-state wavefunction of the qubit Hamiltonian (a task closer to the final energy estimations on a quantum device by updating the covariances with the implemented wavefunction, covering 74 Hamiltonians) and from the CISD wavefunction obtained from PySCF (a task closer to the initial guess and early optimization stages of molecular Hamiltonians, covering 56 Hamiltonians). We decided to use the CISD wavefunction from PySCF rather than diagonalizing a CISD-like subspace in the qubit Hamiltonian, as it allows constructing a solution with the correct spin, since in cases like \ce{H2O} with a stretched bond, the subspace diagonalization yields a quintet state with negligible overlap with the actual solution. For this reason, we limit the set of CISD initializations to neutral, singlet molecules to ensure this is a reasonable approximate wavefunction. Details on the Hamiltonians and full results tables are available in Section III in the SM. \\

We found that the VarSI-G variant generally fails to outperform the SI baseline in both non-overlapping and overlapping settings; see Table \ref{tab:main-spp-comparison} and the SM. For non-overlapping groupings, we find that the refinement variants VarSI-R and VarSI-OR consistently reduce measurement requirements by around 36--39\% across the dataset, regardless of the type of covariances used for scoring and fermion-to-qubit mappings. Considering exact covariances among Hamiltonians without frozen cores, the largest non-overlapping measurement-cost reductions from VarSI-R are observed for \ce{OH-} under the BK mapping (67.7\%), followed by \ce{BH} and \ce{HF}, which achieve reductions of approximately 67.1\% under both JW and BK mappings. Within this all-electron subset, VarSI-OR achieves its largest improvements for \ce{BH}, reaching 79.5\% with BK and 75.6\% with JW, followed by \ce{NH} (68.6\%, BK) and \ce{HF} (63.7\%, JW). Importantly, these improvements largely persist when covariance dictionaries are constructed from approximate CISD wavefunctions. For the 56 matching Hamiltonian--mapping configurations, including those with and without the frozen-core approximation, the mean reductions from VarSI-R are 39.5\% and 38.4\% using exact and CISD covariances, respectively, while VarSI-OR achieves corresponding reductions of 38.4\% and 36.3\%. For the following all-electron examples, \ce{BH} under BK retains reductions of 67.5\% (VarSI-R) and 73.4\% (VarSI-OR) using CISD covariances, compared with 67.1\% and 79.5\% using exact covariances. Similarly, \ce{H2O} under BK achieves reductions of 56.8\% and 54.8\% using CISD covariances, compared with 56.6\% and 57.0\% using exact covariances. For \ce{NH3} (BK) and \ce{HF} (JW), the VarSI-OR reductions relative to the corresponding SI baselines are lower with CISD covariance dictionaries by 19.6 and 11.8 percentage points, respectively, decreasing from 50.1\% to 30.5\% for \ce{NH3} and from 63.7\% to 51.9\% for \ce{HF}. These results show that variance-informed non-overlapping grouping retains much of its effectiveness when covariance estimates come from approximate wavefunctions, although sensitivity to covariance accuracy varies across molecular systems.

\begin{figure}[t!]
    \centering
    \includegraphics[width=\linewidth]{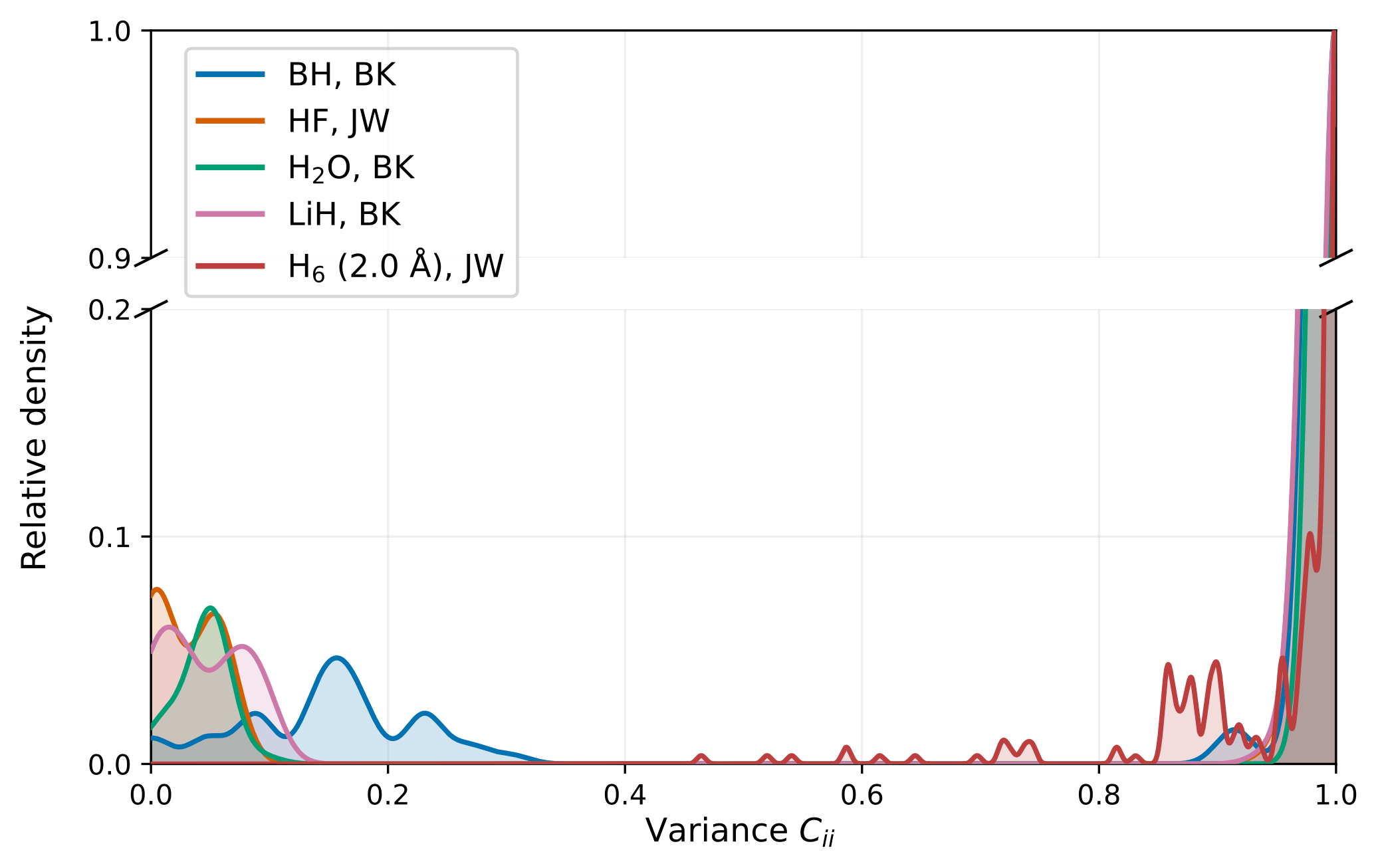}
    \caption{Distribution of the exact FCI variances $C_{ii}$ of the nonidentity Pauli terms in the all-electron STO-3G qubit Hamiltonians for \ce{BH} (BK), \ce{HF} (JW), \ce{H2O} (BK), \ce{LiH} (BK), and linear \ce{H6} at $2.0\,\text{\AA}$ nearest-neighbor spacing (JW). Each density is independently normalized to a maximum of unity, allowing comparison of distribution shapes. We obtained the smoothed distributions using Gaussian kernel density estimation with a Scott bandwidth of $0.2$.
    }
    \label{fig:variances}
\end{figure}

The variance distributions provide further insight into the effectiveness of the VarSI variance-informed grouping. SI uses the uniform-sphere approximation, where the Pauli variances are $C_{ii}=1-\frac{1}{2^{n_q}+1}$, and the covariances satisfy $C_{ij}=0$ for $i\neq j$ \cite{Crawford2021SortedInsertion}. As shown in Fig.~\ref{fig:variances}, \ce{BH} (BK), \ce{HF} (JW), \ce{H2O} (BK), and \ce{LiH} (BK) exhibit substantial populations of low-variance terms, deviating from this approximation and achieving measurement-cost reductions of 79.5\%, 67.1\%, 57.0\%, and 51.5\% relative to SI, respectively. In contrast, \ce{H6} at 2.0~\AA{} (JW) presents the smallest deviations from the idealized scenario that SI targets, reaching a smaller reduction of 22.2\%. A similar qualitative contrast persists after ICS, with the best VarSI initializations yielding reductions of 16.1--45.1\% relative to SI-ICS for the same \ce{BH}, \ce{HF}, \ce{H2O}, and \ce{LiH} configurations, compared with 8.2\% for \ce{H6} at 2.0~\AA{} (JW). Overall, these observations suggest that deviations from the idealized variance description underlying SI offer greater opportunities for variance-informed grouping, consistent with previous reports \cite{huidobromeezs2025discreteflowbasedgenerativemodels}, with the magnitude of the improvement depending on how these deviations contribute to measurement cost, making variance-informed heuristics like VarSI an area of opportunity to exploit these features.

For overlapping ICS measurement initialization, we find that when using covariance dictionaries obtained from the exact ground-state wavefunction, VarSI-O-ICS/VarSI-R-ICS/VarSI-OR-ICS obtain a reduction with respect to SI-ICS of 6.2/10.0/10.3\%, respectively, with maximum reductions of 37.5/38.7/40.8\% for JW mapped Hamiltonians. For BK mappings, the average reductions are 7.1/8.9/11.8\% with max reductions of 37.0/31.6/45.1\%. When initializing the algorithms with CISD covariances, JW-mapped Hamiltonians achieve average reductions of 5.7/8.8/11.4\% and a maximum reduction of 36.4/30.9/48.1\%; BK-mapped Hamiltonians achieve average reductions of 9.8/10.9/13.5\% and a maximum reduction of 36.5/31.4/41.6\%. All refinement variants use 100 sweep steps, as Table \ref{tab:main-spp-comparison} shows no significant improvement with 500 sweep steps. Our results demonstrate that the refinement step is sufficient to significantly improve the measurement requirements and, when combined with our VarSI-O version, can further reduce measurement costs.

\begin{figure}[t!]
    \centering
    \begin{minipage}[t]{0.45\textwidth}
        \centering
        \begin{tikzpicture}
            \node[anchor=south west, inner sep=0] (img) at (0,0)
                {\includegraphics[width=\linewidth]{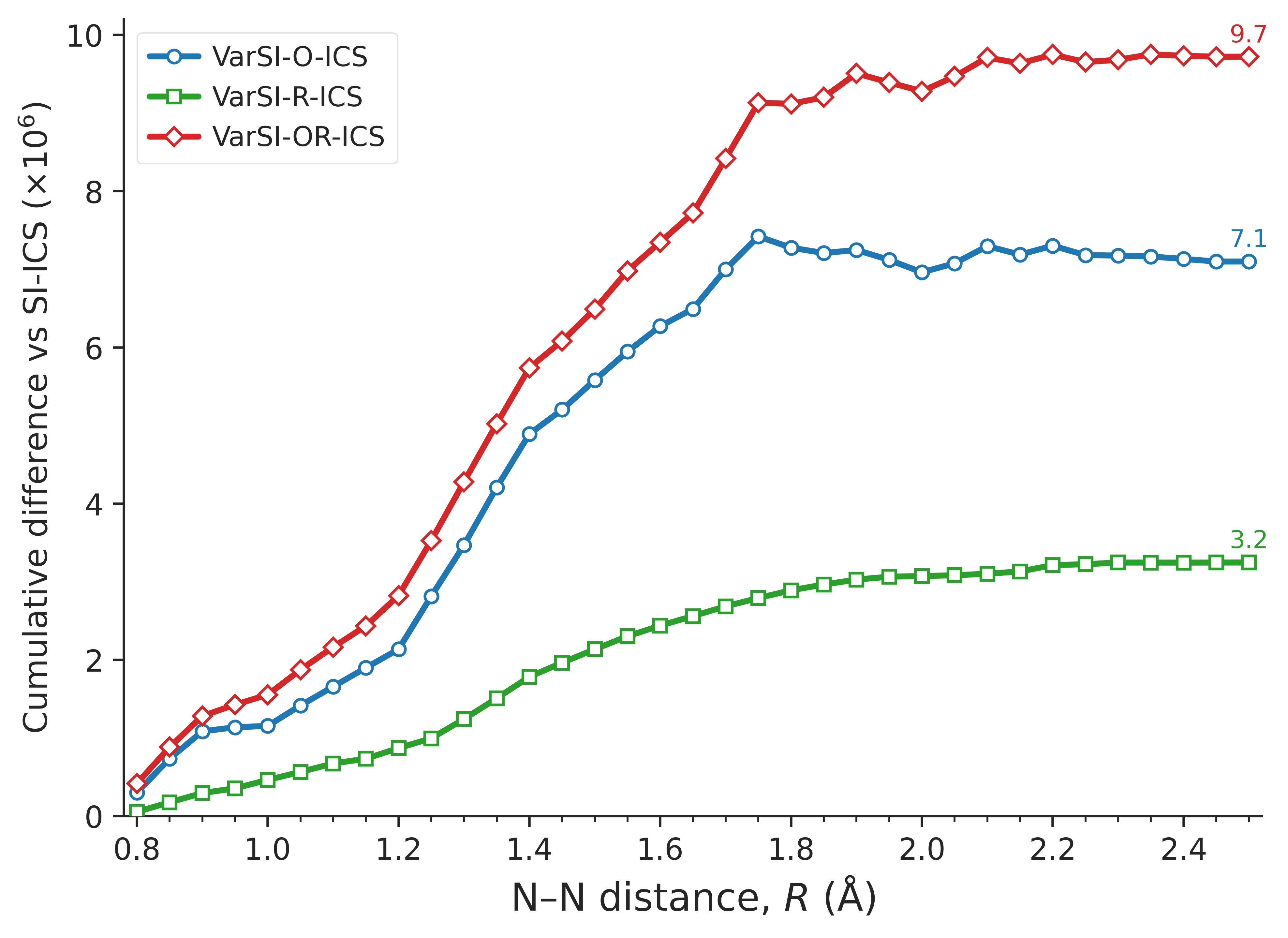}};
            \node[
                anchor=north west,
                fill=white,
                fill opacity=0.85,
                text opacity=1,
                inner sep=0.2pt,
                font=\scriptsize\bfseries
            ] at ([xshift=-4pt,yshift=-5pt]img.north west) {a)};
        \end{tikzpicture}
    \end{minipage}

    \begin{minipage}[t]{0.45\textwidth}
        \centering
        \begin{tikzpicture}
            \node[anchor=south west, inner sep=0] (img) at (0,0)
                {\includegraphics[width=\linewidth]{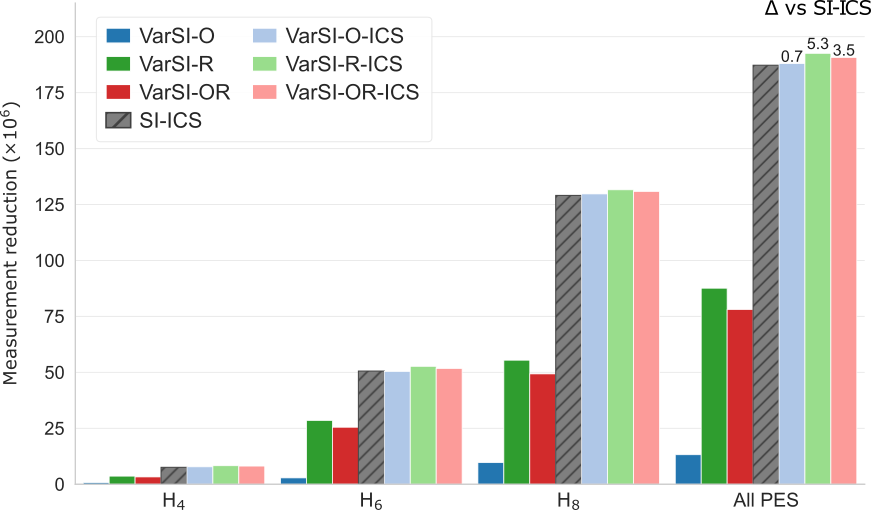}};
            \node[
                anchor=north west,
                fill=white,
                fill opacity=0.85,
                text opacity=1,
                inner sep=0.2pt,
                font=\scriptsize\bfseries
            ] at ([xshift=-4pt,yshift=-5pt]img.north west) {b)};
        \end{tikzpicture}
    \end{minipage}

    \caption{ a) Cumulative measurement improvements over SI-ICS to produce the PES for \ce{N2} dissociation up to a 1 $mE_h$ accuracy using exact covariances. Results shown in millions. Interatomic distance ranges from $R$=0.8--$2.5~\text{\AA}$ using a 0.05 $\text{\AA}$ step. b) Cumulative reductions relative to SI in millions of measurements to reach a $1~\mathrm{m}E_h$ error for PES scans of \ce{H4}, \ce{H6}, \ce{H8} Hamiltonians. Hamiltonians taken from HamLib~\cite{HamLib} over the range $R$=0.5--$2.0~\text{\AA}$ with the JW mapping and a $0.1\text{\AA}$ step. All PES refer to the sum of the strongly correlated systems. The last three columns show, as insets, the differences with respect to SI-ICS for the proposed initializations of the overall task. VarSI-R/VarSI-OR used 100 sweeps for both tasks.}
    \label{fig:RedVarSI}
\end{figure}

To assess our methods on strongly correlated systems and to simulate scenarios that require multiple energy estimates, as is common in quantum chemistry pipelines, we performed two additional tasks. First, we studied the measurement requirements to produce the PES for the \ce{N2} molecule, where multiple bonds are being broken, using interatomic distances in the range $R$=0.8--$2.5~\text{\AA}$ and a 0.05 $\text{\AA}$ step with the STO-3G basis set and the JW mapping. Fig. \ref{fig:RedVarSI}-a) shows the cumulative measurement improvements of the VarSI-ICS initializations against the SI-ICS standard method. For generating the full PES, VarSI-O-ICS, VarSI-R-ICS, and VarSI-OR-ICS reduce the number of measurements required to reach $1~\mathrm{m}E_h$ accuracy by 7.1, 3.2, and 9.7 million, respectively, with VarSI-OR-ICS performing best. These reductions, assuming a repetition delay of $250~\mu s$ and a circuit duration of $100~\mu s$ as reference \cite{DelayDevices,IBMQuantumWorkloadUsage}, translate to approximately 1 hour less of QPU runtime for VarSI-OR-ICS, with $\approx 40$ and $\approx 20$ fewer minutes for VarSI-O-ICS and VarSI-R-ICS, respectively. Similar results are obtained when using approximate covariances from CISD to guide grouping, with reductions in PES construction of 12.1, 3.0, and 14.2 million measurements relative to SI-ICS with the same covariance information, as shown in the SM.

Second, we used the Hamiltonians for \ce{H4}, \ce{H6}, and \ce{H8} from the \texttt{hydrogen\_data} dataset in \texttt{HamLib}, which uses the STO-6G basis set. We explore the measurement cost of building all the potential energy surfaces (PESs) for the symmetric dissociation of the hydrogen chains with a step of $0.1\text{ \AA}$, using SI-ICS, VarSI-R-ICS, VarSI-O-ICS, and VarSI-OR-ICS, going from $R$=0.5--$2.0~\text{\AA}$. For these cases, multiple bonds are being broken simultaneously as well. We considered the JW, BK, and parity mappings. This task comprises 48 energy estimates per mapping (16 per molecule, 144 total), a small number compared to the many energy estimates required in energy optimizations; however, it is useful for simulating a more complex chemistry pipeline with strongly correlated systems. Fig. \ref{fig:RedVarSI}-b) shows the overall measurement reductions to reach $1~\mathrm{m}E_h$ accuracy for constructing the potential energy surface (PES) using the JW mapping for the individual PES and for the full task. We find that the refinement variants VarSI-R and VarSI-OR reduce the total number of required measurements across all PESs by 5.3 and 3.5 million, respectively, relative to SI-ICS. In this case, the reductions amount to $\approx 30$ and $\approx 20$ fewer minutes of QPU runtime for VarSI-R and VarSI-OR. We observe similar results for the BK and parity mappings, regardless of whether the exact ground-state or CISD wavefunctions are used to construct the covariance dictionary, as shown in Tables IV and V of the SM. Overall, we find that SI-ICS was strictly better than all the VarSI-ICS versions in only 5 of the 418 instances tested, all of which used approximate CISD variances, demonstrating the consistent improvements achieved by our initializations.

\section{Conclusions}
\label{sec:conclusions}
In this work, we showed that the proposed variance-informed algorithms for generating non-overlapping groups can substantially reduce the number of measurements required to achieve chemical accuracy across a series of molecular Hamiltonian benchmarks and qubit mappings. The proposed algorithms are simple and readily implementable in current quantum computing packages with electronic-structure functionality, such as Tequila \cite{Kottmann2021Tequila} and OpenFermion \cite{McClean2020OpenFermion}, as well as QML packages like PennyLane \cite{pennylane}, an example is included in the accompanying repository. Variance-guided extensions of largest-degree-first coloring were previously investigated by J{\o}rgensen et al. \cite{jorgensen2026variancereductionmethodsestimation} through the LDVF and LVF heuristics, which select Pauli-group assignments by minimizing the variance of a randomized Horvitz–Thompson estimator with coefficient-norm-based group sampling. VarSI employs covariance-informed insertion updates to directly minimize the measurement requirement under optimal deterministic shot allocation. The proposed methods also incorporate single-term variance ordering and local relocation refinements, and evaluate the resulting non-overlapping partitions as initializations for iterative coefficient splitting. In this direction, our variance distribution analysis suggests that larger improvements in variance-informed algorithms generally correspond to greater deviations from the idealized variances underlying SI. Our results show that the initial grouping used in qubit-based overlapping techniques remains a consequential step in the measurement-reduction pipeline, where additional gains are still possible. Relative to SI baselines, our non-overlapping groupings VarSI-O, VarSI-R, and VarSI-OR reduce measurement costs, with VarSI-R and VarSI-OR emerging as the most promising variants, achieving average reductions of approximately 36--39\%. Importantly, even refinements to the SI groupings used in VarSI-R yield appreciable reductions in measurement costs, providing a straightforward path to resource savings.

We further showed that these improved non-overlapping groupings can serve as effective initializations for overlapping measurement strategies. When used to initialize the ICS overlapping method, the proposed VarSI-OR groupings reduce measurement costs by an average of 10.3--13.5\% relative to SI-ICS, depending on the mapping and covariance dictionaries employed, with maximum reductions of up to 48\%. This allows ICS, when combined with improved initial groupings, to better compete with more expensive frameworks such as SPP. Importantly, the reductions found with VarSI amount to up to 1 hour of QPU time saved in common quantum chemistry applications.
As we mentioned, the classical overhead of producing the necessary covariance dictionaries can be mitigated by using high-quality wavefunctions generated using machine-learning techniques \cite{Mole,ML1RDM,b2026ML2RDM,FlowVQE}. We also anticipate that these improved initializations may be useful for warming up generative policies designed for multiobjective measurement reduction \cite{huidobromeezs2025discreteflowbasedgenerativemodels}. More broadly, we expect that the non-overlapping groupings produced by the algorithms presented here can be used as initializations for other overlapping measurement techniques in qubit space, such as SPP \cite{Choi2022Ghost} and the recently proposed repacking strategy \cite{rowland2026overlappedgroupingsRepacking}.

\section*{Code availability}
The code necessary to reproduce our results, including a parallel covariance dictionary builder, is available at \href{https://github.com/ishume94/VarSI}{https://github.com/ishume94/VarSI}. 

\section*{Acknowledgments}
ILHM acknowledges support from Sandbox AQ's 2025 Research Excellence Scholarship.
This research was enabled by support from the Digital Research Alliance of Canada and NSERC Discovery Grant No. RGPIN-2024-06594. This research was partly enabled by Compute Ontario (computeontario.ca) and the Digital Research Alliance of Canada (alliancecan.ca) support. 

\bibliographystyle{apsrev4-2}
\bibliography{references}

\newpage

\end{document}


\title{Supplementary Material: ``Reducing quantum measurements in qubit-based overlapping grouping methods for quantum energy estimation through better initializations''}

\author{Isaac L. Huidobro-Meezs}
\email{huidobri@mcmaster.ca}
\affiliation{Department of Chemistry and Chemical Biology, McMaster University,\\ Hamilton, ON L8S 4M1, Canada}

\author{Rodrigo A. Vargas-Hern\'andez}
\email{vargashr@mcmaster.ca}
\affiliation{Department of Chemistry and Chemical Biology, McMaster University,\\ Hamilton, ON L8S 4M1, Canada}
\affiliation{Brockhouse Institute for Materials Research, McMaster University,\\ Hamilton, ON, Canada}

\begin{abstract}
The purpose of this supplemental material is to provide the computational cost of the algorithms presented in the main paper and the corresponding pseudocode in Sections \ref{sec:cost} and \ref{sec:pseudo}, respectively, as well as the full results in Section \ref{sec:AddResults} for the algorithms detailed in the main draft. We cover 74 molecular Hamiltonians generated using the STO-3G basis set with Tequila \cite{Kottmann2021Tequila} and PySCF \cite{PySCF} with either the Jordan-Wigner (JW) or Bravyi-Kitaev (BK) mappings. For those 74 Hamiltonians, the covariance dictionary is constructed from the ground-state wavefunction. To evaluate the algorithm's performance with an approximate CISD wavefunction, we selected the 56 Hamiltonians corresponding to neutral molecules with a singlet ground state. To test performance in strongly correlated systems, we look at the dissociation of \ce{H4}, \ce{H6}, and \ce{H8} with the minimal basis STO-6G with the Hamiltonians taken from Ref. \cite{HamLib}, employing both the ground-state and the CISD wavefunction for covariance dictionaries. These Hamiltonians include JW, BK, and Parity mappings and cover interatomic distances from $0.5$--$2~\text{\AA}$  with a $0.1~\text{\AA}$ step, leading to 144 Hamiltonians. In all cases, CISD results are obtained using the CISD covariance dictionary for the scoring function during group generation in our algorithm, and the reported results correspond to $\epsM$ obtained with the ground-state wavefunction for the given groups. Finally, we test the \ce{N2} PES costs using CISD covariances for grouping. Across 418 instances combining results using exact ground-state and CISD covariances, we found that initializing ICS with the standard SI was better than the VarSI initializations in only 5 cases, all of them when using CISD covariances.
\end{abstract}

\maketitle
\newpage
\section{Computational cost}
\label{sec:cost}

Let $N=N_P$ be the number of non-identity Pauli words, $n_q$ the number of qubits, $L$ the number of groups present during a run, and $T_c$ the cost of one pairwise compatibility test.  In a direct binary representation $T_c=O(n_q)$.  Covariance lookup is assumed to be $O(1)$ after the covariance dictionary has been constructed. Constructing the covariance dictionary requires scanning $O(N^2)$ Pauli-word pairs.  In the implementation used here, a covariance entry is evaluated and stored only for commuting pairs, giving a preprocessing cost $O(N^2T_c+N_{\rm comm}C_{\rm cov})$ and memory $O(N_{\rm comm})$, where $N_{\rm comm}\leq N(N+1)/2$ and $C_{\rm cov}$ is the state-dependent cost of evaluating one Pauli covariance.

\subsection{Sorted insertion (SI)}

SI sorts the terms by $|c_i|$, costing $O(N\log N)$, and then checks existing groups until a compatible destination is found.  In the worst case, a new term can be compared with all previously inserted terms before placement. Therefore,
\begin{equation}
    T_\mathrm{SI}=O(N\log N+N^2T_c).
\end{equation}
This is consistent with the $O(n_qN^2)$ worst-case scaling reported for sorted insertion when compatibility tests are performed directly on $n_q$-qubit Pauli words \cite{Crawford2021SortedInsertion}.

\subsection{Ordered VarSI (VarSI-O)}

VarSI-O sorts by $v_i=c_i^2C_{ii}$, also costing $O(N\log N)$ after diagonal covariances are available.  For each term, it scans all compatible existing groups rather than stopping at the first compatible group.  Over the full run, compatibility checks compare each candidate with at most all previously inserted terms, resulting in $O(N^2T_c)$ time.  The variance update in Eq. 9 of the main text requires a cross-covariance sum over the destination group.  Evaluated directly, the total number of covariance lookups is $O(N^2)$.  Hence
\begin{equation}
    T_\mathrm{VarSI-O}=O(N\log N+N^2T_c+N^2).
\end{equation}
With precomputed covariances, VarSI-O has the same quadratic asymptotic scaling in $N$ as SI, but with a larger constant because all compatible destinations are scored.

\subsection{Global greedy VarSI (VarSI-G)}

At an intermediate stage with $r$ remaining terms and $N-r$ placed terms, VarSI-G scans all remaining terms against all possible destinations.  The worst-case work at that stage is $O(r(N-r)(T_c+1))$, where the $+1$ accounts for covariance lookups and arithmetic after compatibility has been checked.  Summing over all stages gives
\begin{equation}
    T_\mathrm{VarSI-G}
    =\sum_{r=1}^{N}O\bigl(r(N-r)(T_c+1)\bigr)
    =O\bigl(N^3(T_c+1)\bigr).
\end{equation}
This cubic scaling is the cost of allowing the insertion order itself to be chosen greedily at every step.

\begin{table}[t!]
\caption{Worst-case grouping costs with a precomputed covariance dictionary.  $N$ is the number of Pauli words, here the pairwise compatibility-test cost ($T_c$) is replaced by its scaling $n_q$, and $N_S$ is the number of refinement sweeps.}
\label{tab:costs}
\begin{tabular}{|l|l|}
\hline
Method & Worst-case grouping cost \\
\hline
SI & $O(N\log N+N^2n_q)$ \\
VarSI-O & $O(N\log N+N^2(n_q+1))$ \\
VarSI-G & $O(N^3(n_q+1))$ \\
VarSI-R, naive & $O(N_S(N^2n_q+N^3))$ \\
VarSI-R, cached & $O(N_SN^2(n_q+1))$ \\
\hline
\end{tabular}
\end{table}

\subsection{VarSI refinement (VarSI-R)}

Let $N_S$ be the number of local-search sweeps. In one sweep,
VarSI-R tests admissible one-term relocations $i:G_\alpha\to G_\beta$.
For such a move, the source-side variance update depends only on
the pair $(i,G_\alpha)$,
\begin{equation}
R_{i,G_\alpha}
=
c_i^2C_{ii}
+
2c_i\sum_{j\in G_\alpha\setminus\{i\}}c_jC_{ij},
\label{eq:source-removal-cache}
\end{equation}
with $V_{G_\alpha\setminus\{i\}}=V_{G_\alpha}-R_{i,G_\alpha}$. The destination-side update depends only on $(i,G_\beta)$,
\begin{equation}
A_{i,G_\beta}
=
c_i^2C_{ii}
+
2c_i\sum_{j\in G_\beta}c_jC_{ij}.
\label{eq:destination-addition-cache}
\end{equation}
A direct implementation that recomputes these sums for every candidate
relocation can require $O(N^3)$ covariance arithmetic per sweep in
unbalanced group configurations. Hence,
\begin{equation}
T_{\mathrm{VarSI\text{-}R}}^{\mathrm{naive}}
=
O\!\left[N_S\left(N^2T_c+N^3\right)\right].
\label{eq:varsi-r-naive-scaling}
\end{equation}
If the removal terms $R_{i,G_\alpha}$ and addition terms $A_{i,G_\beta}$ are cached or maintained during the sweep, each candidate relocation can be scored in $O(1)$ covariance arithmetic after $O(N^2)$ preprocessing. The local-search cost then becomes,
\begin{equation}
T_{\mathrm{VarSI\text{-}R}}^{\mathrm{cached}}
=
O\!\left[N_SN^2(T_c+1)\right].
\label{eq:varsi-r-cached-scaling}
\end{equation}

\section{Pseudocode} \label{sec:pseudo}

Here, we present the pseudocodes for VarSI-O, Fig.~\ref{alg:varsi-o}, VarSI-G, Fig.~\ref{alg:varsi-g}, and VarSI-R, Fig.~\ref{alg:varsi-r}.

\begin{figure}[h!]
\setlength{\abovedisplayskip}{2pt}
\setlength{\belowdisplayskip}{2pt}
\setlength{\abovedisplayshortskip}{2pt}
\setlength{\belowdisplayshortskip}{2pt}
\scriptsize
\begin{algorithmic}[1]
\Require Pauli terms $\mathcal{P}=\{(c_i,P_i)\}_{i=1}^{N_P}$, covariance matrix $C$, and compatibility rule FC or QWC.
\Ensure Non-overlapping compatible grouping $\mathcal{G}$.
\Statex For a group $G_\alpha$, let $V_{G_\alpha}$ be its covariance-based variance. For a grouping $\mathcal{G}$, define
\[
S(\mathcal{G})=\sum_{G_\alpha\in\mathcal{G}}\sqrt{V_{G_\alpha}},
\qquad
\epsilon^2 M(\mathcal{G})=S(\mathcal{G})^2 .
\]
\Statex A term $P_i$ is compatible with $G_\alpha$ if it satisfies the chosen FC or QWC condition with every Pauli word in $G_\alpha$.
\Statex Let $\operatorname{insert}(\mathcal{G},i,G_\alpha)$ denote the grouping obtained by inserting term $i$ into $G_\alpha$; if $G_\alpha=\emptyset$, a new singleton group $\{i\}$ is created.
\Procedure{VarSI-O}{$\mathcal{P},C$}
    \State Sort the terms in $\mathcal{P}$ by decreasing $v_i=c_i^2C_{ii}$
    \State $\mathcal{G}\gets\emptyset$
    \ForAll{terms $i$ in the sorted order}
        \If{there is no group $G_\alpha\in\mathcal{G}$ compatible with $P_i$}
            \State $\mathcal{G}\gets\operatorname{insert}(\mathcal{G},i,\emptyset)$
        \Else
            \State $G^\star\gets
            \operatorname*{arg\,min}_{G_\alpha\in\mathcal{G}}
            S\!\left(\operatorname{insert}(\mathcal{G},i,G_\alpha)\right)$
            \Statex \hspace{\algorithmicindent}
            subject to $P_i$ being compatible with $G_\alpha$
            \State $\mathcal{G}\gets\operatorname{insert}(\mathcal{G},i,G^\star)$
        \EndIf
    \EndFor
    \State \Return $\mathcal{G}$
\EndProcedure
\end{algorithmic}
\caption{Pseudocode for VarSI-O. Terms are processed by decreasing single-term variance and inserted into the compatible group that gives the smallest value of the measurement objective. 
The variance of each candidate group is updated using the covariance matrix $C$.}
\label{alg:varsi-o}
\end{figure}

\begin{figure}[h!]
\setlength{\abovedisplayskip}{2pt}
\setlength{\belowdisplayskip}{2pt}
\setlength{\abovedisplayshortskip}{2pt}
\setlength{\belowdisplayshortskip}{2pt}
\scriptsize
\begin{algorithmic}[1]
\Require Pauli terms $\mathcal{P}$, covariance matrix $C$, and compatibility rule FC or QWC.
\Ensure Non-overlapping compatible grouping $\mathcal{G}$.
\Statex Use $S(\mathcal{G})$ and compatibility as defined in Fig.~\ref{alg:varsi-o}.
\Statex Let $\operatorname{insert}(\mathcal{G},i,G_\alpha)$ denote the grouping obtained by inserting term $i$ into $G_\alpha$; if $G_\alpha=\emptyset$, a new singleton group $\{i\}$ is created.
\Procedure{VarSI-G}{$\mathcal{P},C$}
    \State Sort $\mathcal{P}$ by decreasing $v_i=c_i^2C_{ii}$
    \State Seed $\mathcal{G}$ with the largest-variance term
    \State $R\gets$ remaining terms
    \While{$R\neq\emptyset$}
        \ForAll{$i\in R$}
            \State $\mathcal{A}_i\gets\{G_\alpha\in\mathcal{G}:P_i\text{ is compatible with }G_\alpha\}$
            \If{$\mathcal{A}_i=\emptyset$}
                \State $\mathcal{A}_i\gets\{\emptyset\}$
            \EndIf
        \EndFor
        \State $(i^\star,G^\star)\gets
        \operatorname*{arg\,min}_{i\in R,\;G_\alpha\in\mathcal{A}_i}
        S\!\left(\operatorname{insert}(\mathcal{G},i,G_\alpha)\right)$
        \State $\mathcal{G}\gets\operatorname{insert}(\mathcal{G},i^\star,G^\star)$
        \State $R\gets R\setminus\{i^\star\}$
    \EndWhile
    \State \Return $\mathcal{G}$
\EndProcedure
\end{algorithmic}
\caption{Greedy variance-aware sorted insertion (VarSI-G). At each step, both the next Pauli term and its destination group are chosen to minimize the measurement objective.}
\label{alg:varsi-g}
\end{figure}

\begin{figure}[h!]
\scriptsize
\begin{algorithmic}[1]
\Require Initial grouping $\mathcal{G}_0$, covariance matrix $C$, compatibility rule FC or QWC, and maximum number of sweeps $N_S$.
\Ensure Refined non-overlapping compatible grouping $\mathcal{G}$.
\Statex Use $S(\mathcal{G})$ and compatibility as defined in Fig.~\ref{alg:varsi-o}.
\Statex Write $P_i\sim G_\beta$ when $P_i$ is compatible with every Pauli word in $G_\beta$.
\Statex Let $\operatorname{move}(\mathcal{G},i,G_\alpha\to G_\beta)$ denote moving term $i$ from $G_\alpha$ to $G_\beta$ and updating the affected variances using $C$.
\Procedure{VarSI-R}{$\mathcal{G}_0,C,S_{\max}$}
    \State $\mathcal{G}\gets\mathcal{G}_0$
    \For{$s=1,\ldots,N_S$}
        \State $\mathcal{M}\gets
        \{(i,G_\alpha,G_\beta):G_\alpha,G_\beta\in\mathcal{G},
        G_\alpha\neq G_\beta,\ i\in G_\alpha,\ P_i\sim G_\beta\}$
        \If{$\mathcal{M}=\emptyset$}
            \State \textbf{break}
        \EndIf
        \State $(i^\star,G_\alpha^\star,G_\beta^\star)\gets
        \operatorname*{arg\,min}_{(i,G_\alpha,G_\beta)\in\mathcal{M}}
        S\!\left(\operatorname{move}(\mathcal{G},i,G_\alpha\to G_\beta)\right)$
        \If{$S\!\left(\operatorname{move}(\mathcal{G},i^\star,G_\alpha^\star\to G_\beta^\star)\right)
        \geq S(\mathcal{G})$}
            \State \textbf{break}
        \EndIf
        \State $\mathcal{G}\gets
        \operatorname{move}(\mathcal{G},i^\star,G_\alpha^\star\to G_\beta^\star)$
    \EndFor
    \State \Return $\mathcal{G}$
\EndProcedure
\end{algorithmic}
\caption{Pseudocode for Variance local refinement. The procedure searches over admissible one-term relocations and accepts only moves that reduce the measurement objective. VarSI-R uses sorted insertion as $\mathcal{G}_0$; VarSI-OR uses VarSI-O as $\mathcal{G}_0$.}
\label{alg:varsi-r}
\end{figure}
\clearpage
\newpage

\section{Additional results.}\label{sec:AddResults}
\subsection{Molecular Hamiltonians using the ground state wavefunction from the qubit Hamiltonian for covariances}
\begingroup
\def\STF{\fontsize{9pt}{3.35pt}\selectfont}
\STF
\setlength{\tabcolsep}{2.0pt}
\renewcommand{\arraystretch}{0.74}
\setlength{\LTcapwidth}{\textwidth}
\setlength{\LTleft}{0pt}
\setlength{\LTright}{0pt}
\begin{longtable}{@{}>{\STF}l>{\STF}c>{\STF}c>{\STF}r>{\STF}r>{\STF}r>{\STF}r>{\STF}r>{\STF}r>{\STF}r@{\hspace{0.65em}}!{\vrule width 0.45pt}@{\hspace{0.65em}}>{\STF}r>{\STF}r>{\STF}r>{\STF}r>{\STF}r@{}}
\caption{Molecular benchmark of the measurement-cost $\epsM=(\sum_\alpha\sqrt{V_\alpha})^2$ under full commutativity. Covariance estimates and reported variances were evaluated with ground-state wavefunctions. VarSI-R and VarSI-OR use 100 refinement sweeps. STO-3G basis set used to generate the Hamiltonians. The $R$ column gives the bond distance in \AA. The frozen-core column indicates whether a frozen-core approximation was used. Bold and underlined entries indicate the best and second-best displayed values, respectively, within the non-overlapping and overlapping blocks of each row. Signed reductions are computed relative to SI for non-overlapping groupings and relative to SI-ICS for overlapping groupings.}\label{tab:si-sto3g-fci-fc-s100}\\[-0.6ex]
\toprule
& & & & & \multicolumn{5}{>{\STF}c}{Non-overlapping groupings} & \multicolumn{5}{>{\STF}c}{Overlapping ICS methods} \\
\cmidrule(lr){6-10}\cmidrule(l){11-15}
System & $R$ (\AA) & \shortstack{Frozen\\core} & $N_q$ & $N_P$ & SI & \shortstack{VarSI\\-G} & \shortstack{VarSI\\-O} & \shortstack{VarSI\\-R} & \shortstack{VarSI\\-OR} & \shortstack{SI\\-ICS} & \shortstack{VarSI-G\\-ICS} & \shortstack{VarSI-O\\-ICS} & \shortstack{VarSI-R\\-ICS} & \shortstack{VarSI-OR\\-ICS} \\
\midrule
\endfirsthead
\caption[]{Molecular benchmark of $\epsM$ under full commutativity using exact ground-state variances and 100 refinement sweeps (continued).}\\[-0.6ex]
\toprule
& & & & & \multicolumn{5}{>{\STF}c}{Non-overlapping groupings} & \multicolumn{5}{>{\STF}c}{Overlapping ICS methods} \\
\cmidrule(lr){6-10}\cmidrule(l){11-15}
System & $R$ (\AA) & \shortstack{Frozen\\core} & $N_q$ & $N_P$ & SI & \shortstack{VarSI\\-G} & \shortstack{VarSI\\-O} & \shortstack{VarSI\\-R} & \shortstack{VarSI\\-OR} & \shortstack{SI\\-ICS} & \shortstack{VarSI-G\\-ICS} & \shortstack{VarSI-O\\-ICS} & \shortstack{VarSI-R\\-ICS} & \shortstack{VarSI-OR\\-ICS} \\
\midrule
\endhead
\midrule
\multicolumn{15}{>{\STF}r}{Continued on next page}\\
\endfoot
\bottomrule
\endlastfoot

\multicolumn{5}{>{\STF}c}{} & \multicolumn{10}{>{\STF}c}{\textit{Jordan--Wigner}} \\
\cmidrule(lr){6-15}
\ce{H4} & 0.500 & No & 8 & 184 & 1.032 & 1.485 & 1.049 & \textbf{0.628} & \underline{0.660} & 0.678 & 0.818 & \underline{0.538} & 0.543 & \textbf{0.536} \\
\ce{H4} & 1.000 & No & 8 & 184 & 1.015 & 1.851 & 1.132 & \textbf{0.816} & \underline{0.876} & \textbf{0.706} & 0.996 & \underline{0.771} & \textbf{0.706} & \underline{0.771} \\
\ce{H4} & 1.500 & No & 8 & 184 & 1.062 & \underline{0.889} & 0.974 & \textbf{0.874} & \textbf{0.874} & 0.503 & \textbf{0.402} & 0.503 & \underline{0.501} & \underline{0.501} \\
\ce{H4} & 2.000 & No & 8 & 184 & 1.023 & \textbf{0.658} & 0.929 & 0.940 & \underline{0.925} & 0.330 & \textbf{0.302} & 0.314 & 0.330 & \underline{0.314} \\
\ce{BeH+} & 1.343 & Yes & 10 & 275 & 0.513 & 0.899 & 0.460 & \textbf{0.393} & \underline{0.409} & 0.265 & 0.288 & 0.264 & \textbf{0.253} & \underline{0.262} \\
\ce{LiH} & 1.595 & Yes & 10 & 275 & 0.274 & 0.623 & 0.251 & \underline{0.199} & \textbf{0.176} & 0.108 & 0.108 & \underline{0.106} & \textbf{0.105} & 0.106 \\
\ce{BH} & 1.232 & Yes & 10 & 275 & 1.111 & 1.722 & 1.042 & \underline{0.763} & \textbf{0.677} & 0.583 & \textbf{0.372} & 0.486 & 0.530 & \underline{0.418} \\
\end{longtable}
\endgroup

\clearpage
\newpage
\subsection{Molecular Hamiltonians using the CISD wavefunction for covariances}

\begingroup
\def\STF{\fontsize{9pt}{3.35pt}\selectfont}
\STF
\setlength{\tabcolsep}{2.0pt}
\renewcommand{\arraystretch}{0.74}
\setlength{\LTcapwidth}{\textwidth}
\setlength{\LTleft}{0pt}
\setlength{\LTright}{0pt}
\begin{longtable}{@{}>{\STF}l>{\STF}c>{\STF}c>{\STF}r>{\STF}r>{\STF}r>{\STF}r>{\STF}r>{\STF}r>{\STF}r@{\hspace{0.65em}}!{\vrule width 0.45pt}@{\hspace{0.65em}}>{\STF}r>{\STF}r>{\STF}r>{\STF}r>{\STF}r@{}}
\caption{Molecular benchmark of the measurement-cost $\epsM=(\sum_\alpha\sqrt{V_\alpha})^2$ for neutral STO-3G Hamiltonians with singlet ground states under full commutativity. The VarSI groupings were optimized using CISD covariance estimates with 100 refinement sweeps, while the reported values were evaluated using the exact ground-state variances. The $R$ column gives the bond distance in \AA. The frozen-core column indicates whether a frozen-core approximation was used. Bold and underlined entries indicate the best and second-best values, respectively, within each non-overlapping or overlapping block. Signed reductions are computed relative to SI for non-overlapping groupings and relative to SI-ICS for overlapping groupings.}\label{tab:si-fci-cisd-fc-s100}\\[-0.6ex]
\toprule
& & & & & \multicolumn{5}{>{\STF}c}{Non-overlapping groupings} & \multicolumn{5}{>{\STF}c}{Overlapping ICS methods} \\
\cmidrule(lr){6-10}\cmidrule(l){11-15}
System &$R$ ($\mathrm{\AA}$) & \shortstack{Frozen\\core} & $N_q$ & $N_P$ & SI & \shortstack{VarSI\\-G} & \shortstack{VarSI\\-O} & \shortstack{VarSI\\-R} & \shortstack{VarSI\\-OR} & \shortstack{SI\\-ICS} & \shortstack{VarSI-G\\-ICS} & \shortstack{VarSI-O\\-ICS} & \shortstack{VarSI-R\\-ICS} & \shortstack{VarSI-OR\\-ICS} \\
\midrule
\endfirsthead
\caption[]{Molecular benchmark of $\epsM$ under full commutativity using CISD variances and 100 refinement sweeps (continued).}\\[-0.6ex]
\toprule
& & & & & \multicolumn{5}{>{\STF}c}{Non-overlapping groupings} & \multicolumn{5}{>{\STF}c}{Overlapping ICS methods} \\
\cmidrule(lr){6-10}\cmidrule(l){11-15}
System & $R$ ($\mathrm{\AA}$) & \shortstack{Frozen\\core} & $N_q$ & $N_P$ & SI & \shortstack{VarSI\\-G} & \shortstack{VarSI\\-O} & \shortstack{VarSI\\-R} & \shortstack{VarSI\\-OR} & \shortstack{SI\\-ICS} & \shortstack{VarSI-G\\-ICS} & \shortstack{VarSI-O\\-ICS} & \shortstack{VarSI-R\\-ICS} & \shortstack{VarSI-OR\\-ICS} \\
\midrule
\endhead
\midrule
\multicolumn{15}{>{\STF}r}{Continued on next page}\\
\endfoot
\bottomrule
\endlastfoot
\midrule
\multicolumn{5}{>{\STF}c}{} & \multicolumn{10}{>{\STF}c}{\textit{Jordan--Wigner}} \\
\cmidrule(lr){6-15}
\ce{H4} & 0.500 & No & 8 & 184 & 1.032 & 1.485 & 1.037 & \underline{0.628} & \textbf{0.627} & 0.678 & 0.818 & \underline{0.538} & 0.543 & \textbf{0.497} \\
\ce{H4} & 1.000 & No & 8 & 184 & 1.015 & 1.166 & 1.008 & \textbf{0.816} & \underline{0.876} & \underline{0.706} & \textbf{0.674} & 0.772 & 0.706 & 0.772 \\
\ce{H4} & 1.500 & No & 8 & 184 & 1.062 & \underline{0.885} & 0.974 & \textbf{0.874} & \textbf{0.874} & 0.518 & \textbf{0.406} & 0.518 & \underline{0.517} & \underline{0.517} \\
\ce{H4} & 2.000 & No & 8 & 184 & 1.023 & 1.523 & \textbf{0.994} & 1.022 & \underline{0.999} & \underline{0.674} & 0.809 & 0.678 & \textbf{0.674} & 0.679 \\
\ce{LiH} & 1.595 & Yes & 10 & 275 & 0.274 & 0.508 & 0.252 & \underline{0.199} & \textbf{0.178} & 0.108 & 0.113 & 0.107 & \underline{0.105} & \textbf{0.104} \\
\ce{BH} & 1.232 & Yes & 10 & 275 & 1.111 & 0.991 & 1.042 & \underline{0.749} & \textbf{0.686} & 0.626 & \textbf{0.410} & 0.523 & 0.565 & \underline{0.448} \\
\ce{HF} & 0.917 & Yes & 10 & 275 & 0.778 & 3.516 & 0.903 & \underline{0.594} & \textbf{0.548} & 0.494 & 0.410 & \underline{0.314} & 0.482 & \textbf{0.312} \\
\end{longtable}
\endgroup

\clearpage
\newpage

\subsection{ \texttt{Hydrogen\_data} molecular Hamiltonians taken from \texttt{HamLib} using ground state wavefunction from the qubit Hamiltonian for covariances}

\begingroup
\def\STF{\fontsize{9pt}{3.35pt}\selectfont}
\setlength{\tabcolsep}{2.0pt}
\renewcommand{\arraystretch}{0.74}
\setlength\LTleft{0pt}
\setlength\LTright{0pt}
\setlength\LTcapwidth{\textwidth}
\begin{longtable}{@{}>{\STF}l>{\STF}c>{\STF}c>{\STF}c@{\hspace{0.65em}}!{\vrule width 0.45pt}@{\hspace{0.65em}}>{\STF}c>{\STF}c>{\STF}c>{\STF}c>{\STF}c@{\hspace{0.65em}}!{\vrule width 0.45pt}@{\hspace{0.65em}}>{\STF}c>{\STF}c>{\STF}c>{\STF}c>{\STF}c@{}}
\caption{$\epsM=(\sum_\alpha\sqrt{V_\alpha})^2$ for the \ce{H4}, \ce{H6} and \ce{H8} \texttt{Hydrogen\_data} Hamiltonians from HamLib~\cite{HamLib} which employs the STO-6G basis, using FC groups. The VarSI groupings were optimized using exact covariances with 100 refinement sweeps. Bold and underlined entries indicate the best and second-best displayed values, respectively, within each method block. Signed reductions are computed relative to SI for non-overlapping groupings and relative to SI-ICS for overlapping groupings.} \label{tab:si-hamlib-hydrogen-fci-fc-s100-updated}\\
\toprule
\multicolumn{4}{c}{} & \multicolumn{5}{c}{Non-overlapping groupings} & \multicolumn{5}{c}{Overlapping ICS methods} \\
\cmidrule(lr){5-9}\cmidrule(l){10-14}
System & $R$ ($\mathrm{\AA}$) & $N_q$ & $N_P$ & SI & VarSI-G & VarSI-O & VarSI-R & VarSI-OR & \shortstack{SI\\-ICS} & \shortstack{VarSI-G\\-ICS} & \shortstack{VarSI-O\\-ICS} & \shortstack{VarSI-R\\-ICS} & \shortstack{VarSI-OR\\-ICS} \\
\midrule
\endfirsthead
\caption[]{Continued.}\\
\toprule
\multicolumn{4}{c}{} & \multicolumn{5}{c}{Non-overlapping groupings} & \multicolumn{5}{c}{Overlapping ICS methods} \\
\cmidrule(lr){5-9}\cmidrule(l){10-14}
System & $R$ ($\mathrm{\AA}$) & $N_q$ & $N_P$ & SI & VarSI-G & VarSI-O & VarSI-R & VarSI-OR & \shortstack{SI\\-ICS} & \shortstack{VarSI-G\\-ICS} & \shortstack{VarSI-O\\-ICS} & \shortstack{VarSI-R\\-ICS} & \shortstack{VarSI-OR\\-ICS} \\
\midrule
\endhead
\midrule
\multicolumn{14}{r}{Continued on next page}\\
\endfoot
\bottomrule
\endlastfoot
\multicolumn{4}{c}{} & \multicolumn{10}{c}{\textit{Jordan--Wigner}} \\
\cmidrule(lr){5-14}
\multirow{16}{*}{\ce{H4}} & 0.5 & 8 & 184 & 1.041 & 1.511 & 1.096 & \textbf{0.632} & \underline{0.653} & 0.683 & 0.802 & \underline{0.534} & 0.547 & \textbf{0.487} \\
 & 0.6 & 8 & 184 & 1.058 & 1.240 & 1.165 & \textbf{0.646} & \underline{0.839} & 0.649 & \underline{0.612} & 0.838 & \textbf{0.533} & 0.667 \\
 & 0.7 & 8 & 184 & 1.177 & \textbf{0.737} & 1.173 & 0.824 & \underline{0.754} & 0.923 & \textbf{0.487} & 0.692 & 0.759 & \underline{0.582} \\
 & 0.8 & 8 & 184 & 1.052 & 1.309 & 1.163 & \textbf{0.679} & \underline{0.903} & \underline{0.669} & 0.809 & 0.809 & \textbf{0.536} & 0.811 \\
 & 0.9 & 8 & 184 & 1.078 & 1.121 & 0.862 & \textbf{0.695} & \underline{0.722} & 0.692 & \textbf{0.510} & 0.568 & \underline{0.541} & 0.568 \\
 & 1.0 & 8 & 184 & 1.018 & 1.862 & 1.136 & \textbf{0.819} & \underline{0.879} & \underline{0.706} & 1.003 & 0.773 & \textbf{0.706} & 0.773 \\
 & 1.1 & 8 & 184 & 1.044 & 0.999 & 1.045 & \textbf{0.853} & \underline{0.919} & \underline{0.697} & 0.842 & 0.787 & \textbf{0.697} & 0.787 \\
\end{longtable}
\endgroup

\clearpage
\newpage
\subsection{\texttt{Hydrogen\_data} molecular Hamiltonians taken from \texttt{HamLib} using CISD wavefunction for covariances}

\begingroup
\def\STF{\fontsize{9pt}{3.35pt}\selectfont}
\setlength{\tabcolsep}{2.0pt}
\renewcommand{\arraystretch}{0.74}
\setlength\LTleft{0pt}
\setlength\LTright{0pt}
\setlength\LTcapwidth{\textwidth}
\begin{longtable}{@{}>{\STF}l>{\STF}c>{\STF}c>{\STF}c@{\hspace{0.65em}}!{\vrule width 0.45pt}@{\hspace{0.65em}}>{\STF}c>{\STF}c>{\STF}c>{\STF}c>{\STF}c@{\hspace{0.65em}}!{\vrule width 0.45pt}@{\hspace{0.65em}}>{\STF}c>{\STF}c>{\STF}c>{\STF}c>{\STF}c@{}}
\caption{$\epsM=(\sum_\alpha\sqrt{V_\alpha})^2$ for the \ce{H4}, \ce{H6} and \ce{H8} \texttt{Hydrogen\_data} Hamiltonians from HamLib~\cite{HamLib} which employs the STO-6G basis, using FC groups. The VarSI groupings were optimized using CISD covariances with 100 refinement sweeps, while the reported values were evaluated using the exact ground-state variances. Bold and underlined entries indicate the best and second-best displayed values, respectively, within each method block. Signed reductions are computed relative to SI for non-overlapping groupings and relative to SI-ICS for overlapping groupings.} \label{tab:si-hamlib-hydrogen-cisd-fc-s100-updated}\\
\toprule
\multicolumn{4}{c}{} & \multicolumn{5}{c}{Non-overlapping groupings} & \multicolumn{5}{c}{Overlapping ICS methods} \\
\cmidrule(lr){5-9}\cmidrule(l){10-14}
System & $R$ ($\mathrm{\AA}$) & $N_q$ & $N_P$ & SI & VarSI-G & VarSI-O & VarSI-R & VarSI-OR & \shortstack{SI\\-ICS} & \shortstack{VarSI-G\\-ICS} & \shortstack{VarSI-O\\-ICS} & \shortstack{VarSI-R\\-ICS} & \shortstack{VarSI-OR\\-ICS} \\
\midrule
\endfirsthead
\caption[]{Continued.}\\
\toprule
\multicolumn{4}{c}{} & \multicolumn{5}{c}{Non-overlapping groupings} & \multicolumn{5}{c}{Overlapping ICS methods} \\
\cmidrule(lr){5-9}\cmidrule(l){10-14}
System & $R$ ($\mathrm{\AA}$) & $N_q$ & $N_P$ & SI & VarSI-G & VarSI-O & VarSI-R & VarSI-OR & \shortstack{SI\\-ICS} & \shortstack{VarSI-G\\-ICS} & \shortstack{VarSI-O\\-ICS} & \shortstack{VarSI-R\\-ICS} & \shortstack{VarSI-OR\\-ICS} \\
\midrule
\endhead
\midrule
\multicolumn{14}{r}{Continued on next page}\\
\endfoot
\bottomrule
\endlastfoot
\multicolumn{4}{c}{} & \multicolumn{10}{c}{\textit{Jordan--Wigner}} \\
\cmidrule(lr){5-14}
\multirow{16}{*}{\ce{H4}} & 0.5 & 8 & 184 & 1.041 & 1.509 & 1.060 & \textbf{0.632} & \underline{0.664} & 0.683 & 0.824 & \underline{0.542} & 0.547 & \textbf{0.540} \\
 & 0.6 & 8 & 184 & 1.058 & 1.221 & 1.165 & \textbf{0.646} & \underline{0.839} & 0.649 & \underline{0.593} & 0.842 & \textbf{0.533} & 0.670 \\
 & 0.7 & 8 & 184 & 1.177 & \textbf{0.737} & 1.173 & 0.824 & \underline{0.754} & 0.923 & \textbf{0.487} & 0.694 & 0.759 & \underline{0.583} \\
 & 0.8 & 8 & 184 & 1.052 & 0.968 & 1.163 & \textbf{0.678} & \underline{0.903} & 0.674 & \textbf{0.443} & 0.812 & \underline{0.535} & 0.814 \\
 & 0.9 & 8 & 184 & 1.078 & 1.057 & 0.862 & \underline{0.695} & \textbf{0.694} & 0.701 & \underline{0.560} & 0.571 & \textbf{0.542} & 0.569 \\
 & 1.0 & 8 & 184 & 1.018 & 1.104 & 1.136 & \textbf{0.819} & \underline{0.879} & \underline{0.707} & \textbf{0.678} & 0.774 & \underline{0.707} & 0.774 \\
 & 1.1 & 8 & 184 & 1.044 & 1.598 & 1.047 & \textbf{0.853} & \underline{0.921} & \textbf{0.698} & \underline{0.779} & 0.788 & \textbf{0.698} & 0.788 \\
\end{longtable}
\endgroup

\clearpage
\newpage
\subsection{Cumulative improvements for \ce{N2} PES using the CISD covariances.}
\begin{figure}[h!]
    \centering
    \includegraphics[width=0.55\linewidth]{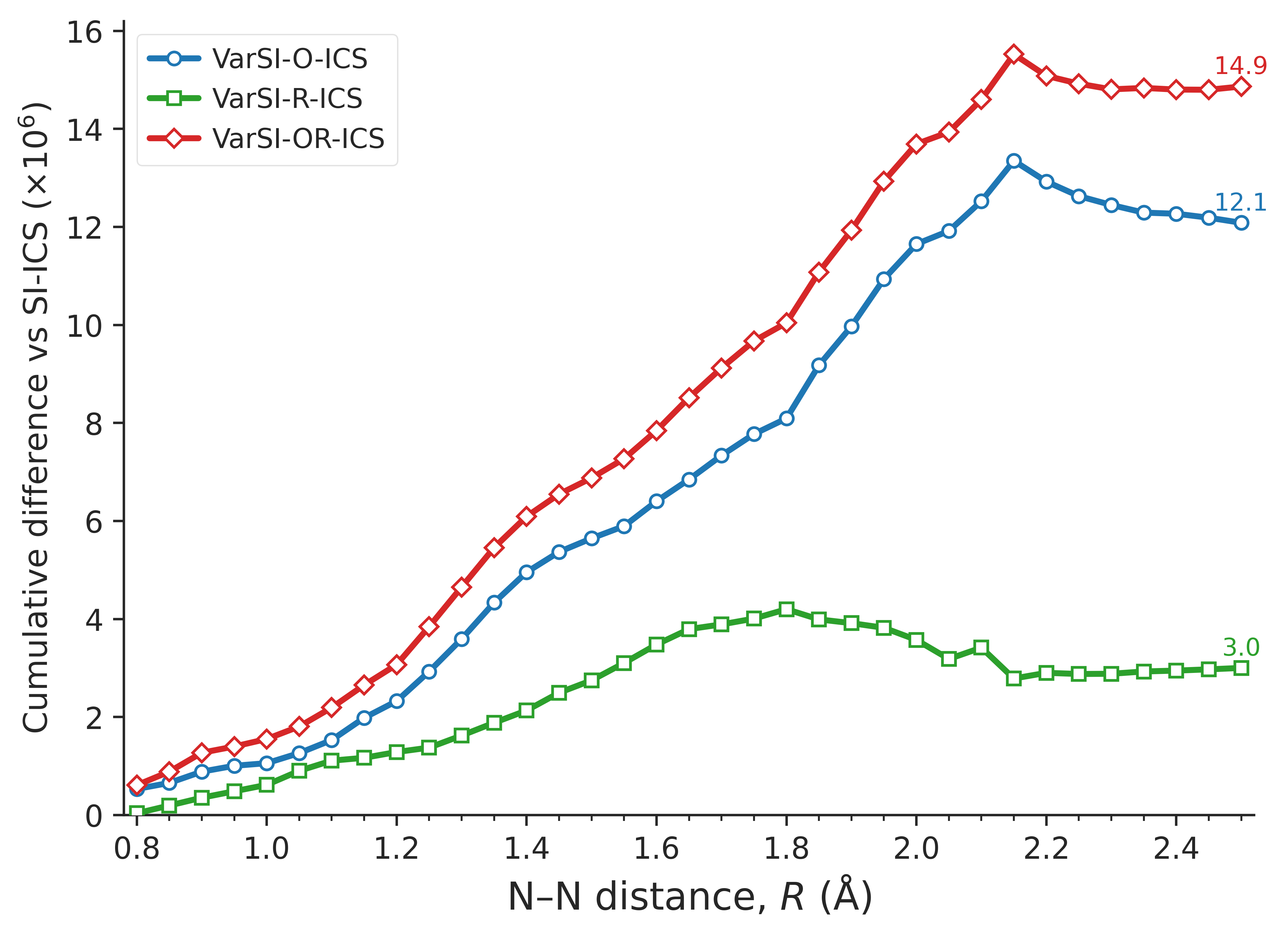}
    \caption{Cumulative measurement improvements over SI-ICS required to produce the PES for \ce{N2} dissociation up to a 1 $mE_h$ accuracy. Results shown in millions. Interatomic distance ranges from $R$=0.8--$2.5~\text{\AA}$ using a 0.05 $\text{\AA}$ step. CISD covariances were used to generate the groupings; the final measurement counts are reported with the ground-state exact variances.}
    \label{fig:placeholder}
\end{figure}

\bibliographystyle{apsrev4-2}
\bibliography{references}